\documentstyle[12pt]{article}

\begin{document}
\makeatletter
\def\siml{\mathrel{\mathpalette\gl@align<}}
\def\simg{\mathrel{\mathpalette\gl@align>}}
\def\gl@align#1#2{\lower.6ex\vbox{\baselineskip\z@skip\lineskip\z@
 \ialign{$\m@th#1\hfill##\hfil$\crcr#2\crcr{\sim}\crcr}}}
\makeatother
\hbadness=10000
\hbadness=10000
\begin{titlepage}
\nopagebreak
\def\thefootnote{\fnsymbol{footnote}}
\begin{flushright}

        {\normalsize
 LMU-TPW 95-8\\
June, 1995   }\\
\end{flushright}
\vspace{1cm}
\begin{center}
\renewcommand{\thefootnote}{\fnsymbol{footnote}}
{\large \bf  Quark Mass Matrices in Orbifold Models}

\vspace{1cm}

{\bf
Tatsuo Kobayashi}
\footnote[1]{Alexander von Humboldt Fellow \\
\phantom{xxx}e-mail: kobayash@lswes8.ls-wess.physik.uni-muenchen.de}

\vspace{1cm}
       Sektion Physik, Universit\"at M\"unchen, \\

       Theresienstr. 37, D-80333 M\"unchen, Germany \\

\end{center}
\vspace{1cm}

\nopagebreak
\begin{abstract}
We study left-right symmetric quark mass matrices whose up- and
down-sectors have the same structure.
This type of realistic mass matrices are derived from orbifold models.

\end{abstract}

\vfill
\end{titlepage}
\pagestyle{plain}
\newpage
\voffset = 0.5 cm


The origin of the fermion masses is one of the most important
problems in particle physics.
Higher dimension couplings could explain the hierarchical structure
of the fermion masses and mixing angles \cite{mass,RRR,IG,BR}.
Underlying symmetries provide selection rules for higher dimension
couplings as well as renormalizable couplings.
In this case entries in mass matrices are written by suppression
factors with powers, which are determined in terms of some types of
quantum numbers under underlying symmetries.
If we assume the left-right symmetry, the quark mass matrices are
dominated by quantum numbers of the quark doublets $Q_i$ as well as
the Higgs fields.
Hence from the assumption of the left-right symmetry it is plausible
to expect that the quark mass matrix of the up-sector
has the same structure as one of the down-sector.
In refs.\cite{IG,BR}, one type of up-down symmetric quark mass
matrices are obtained by using an extra U(1) symmetry.

Superstring theory is the only known candidate for the unified
theory of all the interactions including gravity.
In superstring theory selection rules for higher dimension couplings
are provided by symmetries of a compactified space as well as gauge
symmetries.
In refs.\cite{Nonr,NR}, selection rules for non-renormalizable
couplings are discussed within the framework of orbifold
models \cite{Orbi}.
These selection rules could lead to realistic quark mass matrices.

In this paper we discuss left-right symmetric quark mass matrices
whose up- and down-sectors have the same structure.
We show examples leading to these types of the realistic quark mass
matrices.
The $\mu$-term is also discussed.


In general, the underlying theory like supergravity or superstring
theory includes nonrenormalizable couplings like
$q_{(u,d)j}Q_iH_{2,1}(\theta/M_{2,1})^{n_{ij}}$, where
$q_{(u,d)j}$ denotes the SU(2) singlet quark fields and $H_{2,1}$
are the Higgs fields for the up- and down-sectors.
After the field $\theta$ develops vacuum expectation value (VEV),
this coupling works as the Yukawa coupling with the suppression
factor $\varepsilon_{u,d}^{n_{ij}}=(<\theta>/M_{2,1})^{n_{ij}}$.
That leads to a hierarchical structure in the mass matrices and
the structure depends on underlying symmetries.
In general we can expect $\varepsilon_u \neq \varepsilon_d$.
For example a mixing between light and heavy Higgs fields leads to
$\varepsilon_u \neq \varepsilon_d$ \cite{RRR,IG}.

Here we consider the following left-right symmetric mass matrices:
$$M_{u,d}=d_{u,d}
\pmatrix{
0 & \varepsilon_{u,d}^{n_{12}} & \varepsilon_{u,d}^{n_{13}}  \cr
\varepsilon_{u,d}^{n_{12}}  & \varepsilon_{u,d}^{n_{22}}  &
\varepsilon_{u,d}^{n_{23}}  \cr
\varepsilon_{u,d}^{n_{13}}  & \varepsilon_{u,d}^{n_{23}}  & 1 \cr
}.
\eqno(1) $$
Note that the powers $n_{ij}$ of the up-sector are same as those of
the down-sector.
\renewcommand{\thefootnote}{\fnsymbol{footnote}}
The (1,2), (2,2) and (2,3) elements should be non-vanishing
elements in order to lead to realistic mixing angles \cite{HR}.

At first we consider the case where $\varepsilon_{u,d}^{n_{13}}$ is
negligible.
In this case we have two types of the mass matrices leading to
the geometrical hierarchy $m_3m_1 \approx m_2^2$ as follows,
$$3\ell=n_{12}=3n_{23}, \qquad  n_{22}\geq 2n_{23},
\eqno(2a) $$
$$6\ell=2n_{12}=3n_{22}, \qquad  n_{22}\leq 2n_{23}.
\eqno(2b) $$
We refer to these types, (2a) and (2b), as Type 1 and 2, respectively.
Both types lead to the eigenvalues of masses as $m_3=O(d_{u,d})$,
$m_2=O(\varepsilon_{u,d}^{2\ell} d_{u,d})$ and
$m_1=O(\varepsilon_{u,d}^{4\ell} d_{u,d})$.
In general these types of the quarks mass matrices:
$$M_{u,d}=d_{u,d}
\pmatrix{
0 & a_{u,d}\varepsilon_{u,d}^{3\ell} & 0  \cr
a_{u,d}\varepsilon_{u,d}^{3\ell}  & b_{u,d}\varepsilon_{u,d}^{m}  &
c_{u,d}\varepsilon_{u,d}^{n}  \cr
0  & c_{u,d}\varepsilon_{u,d}^{n}  & 1 \cr
},
\eqno(3) $$
lead to the following CKM matrix:
$$V_{\rm CKM} \approx
\pmatrix{
1-{1 \over 2}(a_d\varepsilon_{d}^{3\ell}/b'_d)^2 &
a_d\varepsilon_{d}^{3\ell}/b'_d &
-a_uc_d\varepsilon_u^{3\ell}\varepsilon_d^n/b'_u \cr
-a_d\varepsilon_{d}^{3\ell}/b'_d  &
1-{1 \over 2}(a_d\varepsilon_{d}^{3\ell}/b'_d)^2
-{1 \over 2}(c_d\varepsilon_d^n)^2   & c_d\varepsilon_d^n \cr
a_dc_d\varepsilon_d^{3\ell}\varepsilon_d^n/b'_d  &
-c_d\varepsilon_d^n & 1-{1 \over 2}(c_d\varepsilon_d^n)^2  \cr
},
\eqno(4) $$
where
$b'_{u,d}=b_{u,d}\varepsilon_{u,d}^m-c_{u,d}^2\varepsilon_{u,d}^{2n}$.
We use the relation $\varepsilon_u<\varepsilon_d$ to get (4).
Eq.(3) with $n=\ell$  corresponds to Type 1, while Type 2 is
obtained by $m=2\ell$.
These types correspond to generalization of the Fritzsch ansatz
\cite{Fritzsch}.
Some phenomenological aspects of generic forms
without the (1,1) and (1,3) elements are discussed.\footnote[1]{
See e.g. ref.\cite{FX}.}
The presence of the (1,3) element does not change the eigenvalues of
the masses if $n_{13} \simg  n_{23}$ for Type 1 and
$n_{13}\simg n_{22}$ for Type 2.
We shall not discuss the case where the (1,3) element is dominant.

The orbifold construction is one of the simplest and most interesting
constructions to derive four-dimensional string vacua \cite{Orbi}.
In orbifold models, string states consist of the bosonic string on
the four-dimensional space-time and a six-dimensional orbifold,
their right-moving superpartners and left-moving gauge parts.
The right-moving fermionic parts are bosonized and momenta of
bosonized fields span an SO(10) lattice.
An orbifold is obtained through a division of a six-dimensional space
$R^6$ by a six-dimensional lattice and its automorphism $\theta$.
Closed strings on the orbifold are classified into untwisted and
twisted sectors.
For the $\theta^k$-twisted sector $T_k$, the string coordinate
satisfies the following boundary condition:
$$
 x_\nu(\sigma=2 \pi)=\theta^kx_\nu(\sigma=0)+e_\nu,
\eqno(5)$$
where $e_\nu$ is a lattice vector.
A zero-mode of this string satisfies the same condition as (5) and
it is called a fixed point.
The fixed point is represented in terms of the space group element
$(\theta^k,e_\nu)$.
All fixed points in $T_k$ are not fixed under $\theta$.
To obtain $\theta$-eigenstates, we have to take linear combinations
of states corresponding to fixed points \cite{KO1,KO2}.
These linear combinations have eigenvalues
$\gamma =\exp[2 \pi i m/k]$.
We take a complex basis $(X_i,\overline X_i)$ $(i=1 \sim 3)$ for
the compactified space, e.g. $X_1=x_1+ix_2$.
Oscillated states in $T_k$ are created by $\partial X_{i(k)}$ and
$\partial \overline X_{i(k)}$.

Couplings are calculated by using vertex operators $V_a$
corresponding to states \cite{FMS,Yukawa}.
Nonvanishing couplings are invariant under a symmetry of each part
of string states.
\renewcommand{\thefootnote}{\fnsymbol{footnote}}
Coupling terms are allowed if they are gauge invariant and space
group invariant. \footnote[2]{
For the selection rule due to the space group, see in detail
ref.\cite{KO2}.}
In addition the SO(10) momentum should be conserved and a product of
eigenvalues $\gamma_a$ should satisfy $\prod_a \gamma_a=1$.
Further the corresponding correlation function $<V_1 \cdots V_n>$
should invariant under a $Z_N$ rotation of oscillators as
$$
\partial X_{i(k)} \rightarrow e^{2 \pi i kv^i} \partial X_{i(k)},
\eqno(6)$$
where $e^{2 \pi i v^i}$ are eigenvalues of $\theta$ in the
complex basis $(X_i,\overline X_i)$.
Note that vertex operators corresponding to non-oscillated states
includes the oscillators except the $-1$ or $-1/2$ picture \cite{FMS}.
These selection rules are discussed explicitly in ref.\cite{NR}.

\renewcommand{\thefootnote}{\fnsymbol{footnote}}
For example, the $Z_6$-II orbifold has eigenvalues $v_i=(2,1,-3)/6$.
For $T_1$, $T_2$ and $T_4$, there are three fixed points on the
first plane.
These fixed points are denoted as $(\theta^k,ie_1)$ with $i=0,1,2$.
Couplings including $T_4$ are allowed in the following forms:
$$T_1^{2\ell}T_4^m,\quad T_3^{2\ell}T_4^{3m},
\quad T_1^{2\ell}T_2^mT_4^n, \quad T_1^{\ell}T_3^mT_2^pT_4^q,
\eqno(7)$$
where $\ell >0,m>0,n>0$ and $p$ or $q>0$.\footnote[3]
{The $Z_6$-II orbifold models allow other couplings.
See in detail ref.\cite{NR}.}
The first coupling is allowed if $\ell+2m=9i+3$ and $2\ell+m=9j+3$.
The third and fourth couplings are forbidden unless they are point
group invariant.
Further the third coupling should satisfy that $\ell =2p+1$ and
$2\ell +2m +n=3i$.
This orbifold construction allows only the $T_1^2T_4$ and $T_1T_2T_3$
couplings as renormalizable couplings of the twisted sectors.

Here we study models to lead to (2.a,b) using the above selection
rules.
We restrict ourselves to the simple cases where $n_{12}=3$, i.e.
$\ell=1$ in (2.a,b) and (3).
In order to obtain a left-right symmetric mass matrix, it is simple
to assign one generation of the left- and right-handed quarks to
the same twisted sector.
We assign the Higgs fields $H_{2,1}$ to $T_4$ and third generations
of the quark fields to $T_1$ so that the renormalizable coupling
$T_1^2T_4$ leads to the (3,3) element.
Only the $T_1T_3T_4^2$ coupling induce the Yukawa coupling with the
factor $\varepsilon$, while the couplings with $\varepsilon^2$ are
generated by $T_1^2T_2T_4^2$, $T_1T_2^2T_3T_4$, $T_1^2T_3^2T_4$ and
$T_3^2T_4^3$ couplings.
In addition $T_1^2T_2^3T_4$, $T_1T_2T_3T_4^3$, $T_1^3T_2T_3T_4$ and
$T_1T_3^3T_4^2$ couplings lead to the Yukawa couplings with
$\varepsilon^3$.
We have to assign the second generation to $T_3$ or $T_4$ to obtain
the (2,3) element of Type 1 as $\varepsilon$.
Further we need the $\theta$ field with the VEV in $T_4$ and $T_3$
if the second generation belongs to $T_3$ and $T_4$, respectively.
In both cases, the (2,2) element is obtained as $\varepsilon^2$
through the $T_3^2T_4^3$ coupling.
If we assign the first generation to $T_1$ for the later case, the
$(T_1T_4^2)T_3^3$ coupling provides the (1,2) element as
$\varepsilon^3$.
In this case we can assign the fixed points of the first and third
generations and the Higgs fields to forbid the (1,1) and/or (1,3)
elements, \footnote[4]{If the fixed points of $T_1$ and $T_4$ on
the first plane do not satisfy the space group invariance, one
cannot generate these elements by the $\theta$ field with the VEV
in $T_3$.}
although without this selection rule due to the space group these
elements are allowed as the renormalizable coupling, $T_1^2T_4$.
Thus we obtain the following mass matrices:
$$M_{u,d}=d_{u,d}
\pmatrix{
0 & \varepsilon_{u,d}^3 & 0  \cr
\varepsilon_{u,d}^3  & \varepsilon_{u,d}^2  & \varepsilon_{u,d}  \cr
0  & \varepsilon_{u,d}  & 1 \cr
}.
\eqno(8) $$
This corresponds to both of the types and is obtained in
refs.\cite{IG,BR} by using an extra U(1) symmetry.
We can obtain the similar mass matrices in the case where the first
and second generations are assigned to $T_1$ and $T_3$, respectively
and $\theta$ fields in $T_3$ and $T_4$ develop VEVs.
In this case it is hard to forbid completely the (1,1) and (1,3)
elements.
Each of these elements is induced through the $(T_1^2T_4)T_3^2T_4^3$
or $(T_1^2T_4)T_4^{9}$ couplings, although this factor is enough
suppressed.
We can easily obtain the mass matrices with the suppressed (1,1) and
(1,3) elements, if we assign the first and third generations to the
same twisted sectors.
It is hard to derive Type 1 with more suppressed (2,2) element, i.e.
$m>2$ in the $Z_6$-II orbifold models, because the (2,2) element is
naturally induced by the $T_3^2T_4^3$ coupling.

Next we discuss examples leading to Type 2 with $n>1$.
It is easy to derive the (2,2) element as $\varepsilon^2$  for any
assignment of the second generation.
However, it is hard to obtain the (1,2) element as $\varepsilon^3$
when the second generation is assigned to $T_1$.
If the second generation belongs to $T_4$, the matrix (8) is
naturally obtained as above.
Thus we consider the case where the second generation is assigned to
$T_3$.
Further we assign the first generation to $T_2$.
Suppose that $\theta$ fields in $T_1$, $T_2$ and $T_4$ develop VEVs.
Then we obtain the (2,2) element as $\varepsilon^2$ through the
$(T_3^2T_4)T_4^2$ coupling.
In this case the (1,1), (1,2) and (2,3) element are provided as
$\varepsilon^3$, $\varepsilon^2$ and $\varepsilon$ by the
$(T_2^2T_4)(T_1^2T_2)$, $(T_2T_3T_4)(T_1T_2)$ and
$(T_1T_3T_4)T_4$ coupling without taking into account other
selection rules.
We assume that the $\theta$ fields with the VEVs in $T_2$ and $T_4$
have $\gamma=-1$, while the other fields have $\gamma=1$.
Then the above couplings are forbidden.
Instead the $(T_2^2T_4)(T_1^6T_4^2)$, $(T_2T_3T_4)T_1^3$ and
$(T_1T_3T_4)T_2^2$ couplings generate the (1,1), (1,2)  and (2,3)
elements as $\varepsilon^8$, $\varepsilon^3$ and $\varepsilon^2$,
respectively.
Then we obtain the following quark mass matrix:
$$M_{u,d}=d_{u,d}
\pmatrix{
\varepsilon_{u,d}^8 & \varepsilon_{u,d}^3 & \varepsilon_{u,d}^3 \cr
\varepsilon_{u,d}^3 & \varepsilon_{u,d}^2 & \varepsilon_{u,d}^2 \cr
\varepsilon_{u,d}^3 & \varepsilon_{u,d}^2 & 1 \cr
},
\eqno(9) $$
where the (1,3) element is obtained through the
$(T_1T_2T_4)(T_1T_2^2)$ coupling.
We can obtain more suppressed (1,3) element if we take into account
the selection rule due to the space group invariance.
For example we assign the fixed point of the $T_1$ field with the
VEV to $(\theta,e_1)$ on the first plane, while the fixed point of
the other fields is assigned to the origin $(\theta^k,0)$.
Then the (1,3) element is derived as $\varepsilon^9$ through the
$(T_1T_2T_4)T_1^9$ coupling, while the $(T_2T_3T_4)T_1^3$ coupling
of the (1,2) element is allowed.
Other assignment of the fixed points can lead to similar mass
matrices.
We neglect the (1,1) and (1,3) elements to obtain the following
mass matrices:
$$M_{u,d}=d_{u,d}
\pmatrix{
0 & \varepsilon_{u,d}^3 & 0 \cr
\varepsilon_{u,d}^3 & \varepsilon_{u,d}^2 & \varepsilon_{u,d}^2 \cr
0 & \varepsilon_{u,d}^2 & 1 \cr
}.
\eqno(10) $$
Eq.(10) is one of the simplest mass matrices \cite{Xing}.
Because four texture zeros are included totally for the up- and
down-sectors and for each matrix there appear only three hierarchies,
1, $\varepsilon^2$ and $\varepsilon^3$.
This type can lead to the realistic mixing angles (4).
In this case the mixing angles satisfy the following relations:
$$|V_{us}|^2 \approx |V_{cb}| \approx {m_d \over m_s}, \quad
{|V_{ub}| \over |V_{cb}|} \approx \sqrt {m_u \over m_c}, \quad
{|V_{td}| \over |V_{ts}|} \approx \sqrt {m_d \over m_s}.
\eqno(11)$$
We can derive realistic mass matrices similar to (9) and (10) in the
case where the second generation is assigned to $T_2$.
In this case we need several fields with VEVs.

We can derive Type 2 with $n>2$.
For example we assign the first and second generations to $T_1$ and
$T_2$, respectively.
Suppose that $\theta$ fields in $T_1$ and $T_3$ develop VEVs.
Then we obtain the same mass matrices as (10) except the (2,3)
element replacing $\varepsilon^3$.
Further we can derive Type 2 with $n>3$, but the larger values of
$n$ do not lead to the realistic mixing angles (4).

We have derived some examples of Type 1 and 2 (2a,b) with $n_{12}=3$.
Similarly we could obtain these types of the quark mass matrices
with $n_{12}=3\ell$ ($\ell >1$), although in some cases it is hard
to rule out lower dimension operators.
It is also interesting to derive the quark mass matrices where the
geometrical hierarchy is obtained approximately, e.g. the case with
$3n_{12}=4n_{22}$ and $n_{23} \geq n_{22}$.
This example corresponds to some down-sectors in ref.\cite{RRR}.

World-sheet instantons lead to another suppression factor as
$\exp[-a_{ij}T]$, where $T$ is the moduli parameter and $a_{ij}$ is
a constant \cite{Yukawa}.
These effects provide the coefficients like (3).
Also CP phases are induced to some elements in the case with
nonvanishing background anti-symmetric tensors \cite{anti}, although
in some orbifold models the physical CP phase do not appear in the
limit $\varepsilon_{u,d} \rightarrow 0$\cite{NR}.\footnote[5]{
See also ref.\cite{CP}.}
The suppression factor $\exp[-a_{ij}T]$ and the CP phase of each element
are in general different from those of other elements.
The difference between $d_u$ and $d_d$ can be explained by the
difference of the VEVs of $H_{2,1}$ or the world-sheet instanton
effects.\footnote[6]
{We could also explain this difference by assuming an extra U(1)
symmetry.}
Further, we have to estimate radiative corrections from a high
energy to a low energy to compare the prediction with experiment.
However, the hierarchical structure, i.e. the powers $n_{ij}$ are
not changed \cite{OP,RRR}, while the coefficients in (3) are
changed by $O(1)$.

At last we discuss the $\mu$-term.
In the above examples, both of the Higgs fields, $H_{2,1}$ are
assigned to $T_4$.
Thus the $\mu$-term is forbidden as a renormalizable couplings.
After the symmetry breaking, some high dimension operators could
generate the effective $\mu$-term, which is naturally suppressed.
For instance, the $\mu$-term in the example leading to (10) should
include $T_1^6T_2^2T_4^2$ coupling.


To sum up, We have studied on left-right symmetric quark mass
matrices, which have the same structure for the up- and down-sectors.
We have shown examples leading to these type of the mass matrices
within the framework of the orbifold models.
For the $Z_6$-II orbifold models, it is hard to obtain Type 1
except (8).
Eq.(10) is one of the simplest and most realistic quark mass
matrices, which cannot be derived by assuming an extra U(1) symmetry.
The above analyses constrain how to assign the quark fields to the
twisted sectors.
That is very useful to model building.

\vspace{0.8 cm}
\leftline{\large \bf Acknowledgement}
\vspace{0.8 cm}

The author would like to thank Z.~Xing for numerous useful
discussions and he also thanks C.S.~Lim, H.~Nakano and
D.~Suematsu.



\end{document}